\documentclass[aps,twocolumn,superscriptaddress,prl]{revtex4}

\usepackage{amssymb}
\usepackage{amsmath}
\usepackage{graphicx}

\def\g{\textsl{\textrm{g}}}
\def\fnod{f_{\mathrm{0}}}
\def\fc{f_{\mathrm{C}}}
\def\Rd{R_{\mathrm{D}}}
\def\Vd{V_{\mathrm{D}}}
\def\squm{\mathrm{\mu m^2}}
\def\Vc{V_{\mathrm{C}}}
\def\Vt{V_{\mathrm{T}}}
\def\Volt{\mathrm{V}}

\def\mT{\mathrm{mT}}
\def\dB{\Delta B}
\def\dVc{\Delta V_{\mathrm{C}}}
\def\dVt{\Delta V_{\mathrm{T}}}
\def\Tesla{\mathrm{T}}
\def\dRd{\delta R_{\mathrm{D}}}
\def\FP{Fabry-P\'{e}rot~}

\def\phinod{\phi_{\mathrm{0}}}
\def\EC{E_{\mathrm{C}}}

\begin{document}
\title{Distinct Signatures For Coulomb Blockade and Aharonov-Bohm Interference in Electronic \FP Interferometers}
\author{Yiming~Zhang}
\affiliation{Department of Physics, Harvard University, Cambridge,
Massachusetts 02138, USA}
\author{D.~T.~McClure}
\affiliation{Department of Physics, Harvard University, Cambridge,
Massachusetts 02138, USA}
\author{E.~M.~Levenson-Falk}
\affiliation{Department of Physics, Harvard University, Cambridge,
Massachusetts 02138, USA}
\author{C.~M.~Marcus}
\affiliation{Department of Physics, Harvard University, Cambridge,
Massachusetts 02138, USA}
\author{L.~N.~Pfeiffer}
\affiliation{Bell Laboratories, Alcatel-Lucent Technologies, Murray
Hill, NJ 07974, USA}
\author{K.~W.~West}
\affiliation{Bell Laboratories, Alcatel-Lucent Technologies, Murray
Hill, NJ 07974, USA}
\date{\today}

\begin{abstract}

Two distinct types of magnetoresistance oscillations are observed in
two electronic \FP interferometers of different sizes in the integer
quantum Hall regime. Measuring these oscillations as a function of
magnetic field and gate voltages, we observe three signatures that
distinguish the two types. The oscillations observed in a
$2.0~\squm$ device are understood to arise from the Coulomb blockade
mechanism, and those observed in an $18~\squm$ device from the
Aharonov-Bohm mechanism. This work clarifies, provides ways to
distinguish, and demonstrates control over, these distinct physical
origins of resistance oscillations seen in electronic \FP
interferometers.

\end{abstract}
\maketitle

Mesoscopic electronics can exhibit wave-like interference
effects~\cite{UCFReview,Liang01,Ji03,Roulleau07}, particle-like charging
effects~\cite{QDReview}, or a complex mix of both \cite{Aleiner02}. Experiments
over the past two decades have investigated the competition between wave and
particle properties~\cite{Buks98}, as well as regimes where they
coexist~\cite{Glattli91,Folk96, Cronenwett97, Aleiner02}. The electronic \FP
interferometer (FPI)--- a planar two-contact quantum dot operating in the quantum
Hall regime---is a system where both interference and Coulomb interactions can
play important roles. This device has attracted particular interest recently due
to predicted signatures of fractional~\cite{Chamon97} and
non-Abelian~\cite{Stern06,Bonderson06,Ilan08} statistics. The interpretation of
experiments, however, is subtle, and must account for the interplay or charging
and interference effects in these coherent confined structures.

Early measurements by van Wees \textit{et~al.}~\cite{vanWees89}
demonstrated resistance oscillations as a function of magnetic field
in an electronic FPI, with an interpretation given in terms of
Aharonov-Bohm (AB) interference of edge states. More recently,
experimental~\cite{Alphenaar92CB,Taylor92,Camino07CB,Godfrey07} and
theoretical~\cite{Dharma92,Rosenow07,Ihnatsenka08} investigations
indicate that Coulomb interaction plays a critical role in these
previously observed conductance oscillations---as a function of both
magnetic field and electrostatic gate voltage---suggesting an
interpretation in terms of field- or gate-controlled Coulomb
blockade (CB). Other recent experiments studying fractional charge
and statistics in FPI's~\cite{caminoFS,Willett08} interpret
resistance oscillations as arising from AB interference while taking
the gate-voltage period as indicating a change of a quantized
charge.

\begin{figure}[b!]
\center \label{fig1}
\includegraphics[width=3.00in]{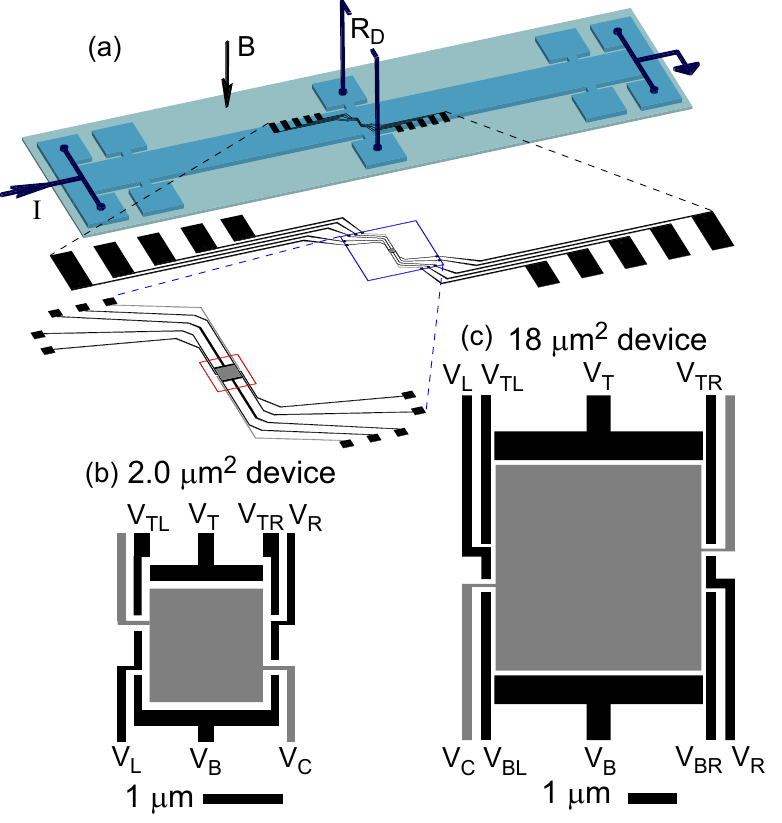}
\caption{\footnotesize{Measurement setup and devices. (a) Diagram of
the wet-etched Hall bar, surface gates, and measurement
configuration. Diagonal resistance, $\Rd$, is measured directly
across the Hall bar, with current bias, $I$. Subsequent zoom-ins of
the surface gates are also shown; the red box encloses the detailed
gate layouts for the device shown in (c). (b,c) Gate layouts for the
$2.0~\squm$ and $18~\squm$ devices, respectively. The areas quoted
refer to those under $\Vc$.}}
\end{figure}

In this Letter, we report oscillations of resistance as a function
of perpendicular magnetic field, $B$, and gate voltage in FPI's of
different sizes. Oscillations in the smaller  ($2.0~\squm$) device
are consistent with the interacting (CB) interpretation, while those
in the larger ($18~\squm$) device are consistent with noninteracting
AB interference. Specifically, three signatures that distinguish the
two types of oscillations are presented: The magnetic field period
is roughly proportional to $B$ for CB, but field-independent for AB;
The gate-voltage period is field-independent for CB, but
proportional to $1/B$ for AB;  Resistance stripes in the
two-dimensional plane of $B$ and gate voltage have a positive
(negative) slope in the CB (AB) regime.

The devices were fabricated on a high-mobility two-dimensional
electron gas (2DEG) residing in a 30~nm wide GaAs/AlGaAs quantum
well 200~nm below the chip surface, with Si $\delta$-doping layers
100~nm below and above the quantum well. The mobility is $\sim2,000
~\mathrm{m^2/Vs}$ measured in the dark, and the density is $2.6
\times 10^{15}~\mathrm{m}^{-2}$. Surface gates that define the FPI's
are patterned using electron-beam lithography on wet-etched Hall
bars [see Fig.~1(a)]. These gates come in from top left and bottom
right, converging near the middle of the Hall bar. Figures~1(b) and
1(c) show gate layouts for the $2.0~\squm$ and $18~\squm$
interferometers. All gate voltages except $\Vc$ are set around $\sim
-3~\Volt$ (depletion occurs at $\sim -1.6~\Volt$). Voltages, $\Vc$,
on the center gates are set near $0~\Volt$ to allow fine tuning of
density and area.

Measurements are made using a current bias $I=400~\mathrm{pA}$, with
$B$ oriented into the 2DEG plane as shown in Fig.~1(a). The diagonal
resistance, $\Rd \equiv d\Vd/dI$ is related to the dimensionless
conductance of the device $\g = (h / e^2) / \Rd$~\cite{Miller07}.
Here, $\Vd$ is the voltage difference between edge states entering
from the top right and bottom left of the device. Figure~2(a) shows
$\Rd$ as a function of $B$, displaying several quantized integer
plateaus. Figures~2(b) and 2(c) show the zoom-ins below the $\g = 1$
and $2$ plateaus, respectively, displaying oscillations in $\Rd$ as
a function of $B$, with periods $\dB = 2.1~\mT$ and $1.1~\mT$. This
$\dB$ of $2.1~\mT$ corresponds to one flux quantum, $\phinod \equiv
h / e$, through an area $A=2.0~\squm$; hence $1.1~\mT$ corresponds
to $\phinod / 2$ through about the same area. Figure~2(d) shows
$\dB$, measured wherever oscillations appear, as a function of $B$;
a linear fit constrained to pass through the origin shows that $\dB$
is almost proportional to $B$. Zoom-ins of the data in Fig.~2(d)
below the $\g = 1$ and $2$ plateaus are shown in Figs.~2(e) and 2(f)
and clearly show that for both cases, the data are flatter than the
linear fit.

\begin{figure}[!]
\center \label{fig2}
\includegraphics[width=3.15in]{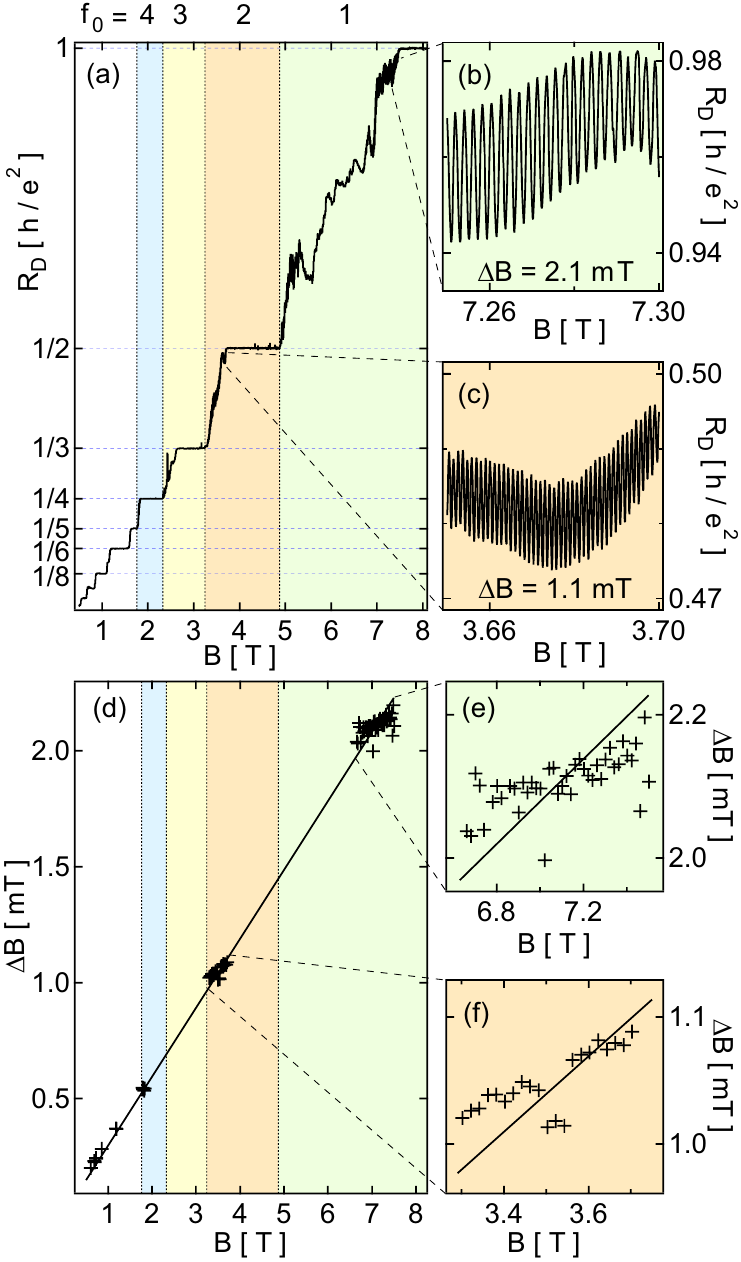}
\caption{\footnotesize{Oscillations in $\Rd$ as a function of
magnetic field, $B$, for the $2.0~\squm$ device. (a) $\Rd$ as a
function of $B$, showing well-quantized integer plateaus. Different
colored backgrounds indicate different numbers of fully-occupied
LL's, $\fnod$, through the device. (b, c) Zoom-ins of the data in
(a), at $\fnod=1$ and 2, respectively, showing oscillations in
$\Rd$, and their $B$ periods, $\dB$. (d) Observed $\dB$ as a
function of $B$, with a straight-line fit through the origin. (e, f)
Zoom-ins of the data in (d) at $\fnod=1$ and 2, respectively.}}
\end{figure}

This approximate proportionality between $\dB$ and $B$ is
inconsistent with simple AB oscillations, which would give a
constant $\dB$ corresponding to one flux quantum through the area of
the device. However, a recent theoretical analysis that accounts for
Coulomb interaction between edge states found that for $\fc$
occupied Landau levels (LL's) in the two constrictions, $\dB =
(\phinod / A) / \fc$ for weak forward tunneling of the $(\fc +
1)^\mathrm{th}$ level, and $\dB = ( \phinod / A ) / (\fc - 1)$ for
weak backscattering of the $\fc^\mathrm{th}$ level \cite{Rosenow07}.
Interpolating between these two limits, we expect when the device
conductance, $\g$, is anywhere between $\fnod$ and $\fnod + 1$, that
$\dB = (\phinod / A) / \fnod$. Here, $\fnod$ is the number of fully
occupied LL's passing through the device (represented by different
colored backgrounds in Fig.~2 for $\fnod$ = 1 to 4). Note that this
model requires the $(\fnod + 1)^\mathrm{th}$ LL to be partially
filled in both constrictions; otherwise, no oscillations are
expected.

In this picture, on the riser of $\Rd$ where $\fnod < \g < \fnod + 1$, the $(\fnod
+ 1)^\mathrm{th}$ and higher LL's will form a quasi-isolated island inside the
device that will give rise to Coulomb blockade effects for sufficiently high field
and large charging energy,
\begin{equation*}
\EC = \frac{e^2}{2C} (\fnod \cdot B A / \phinod + N - \alpha
V_{\mathrm{gate}})^2,
\end{equation*}
 where $N$ is the number of electrons on the island, $C$ is the total
capacitance, and $\alpha$ is the lever-arm associated with gate
voltage $V_{\mathrm{gate}}$ \cite{Rosenow07}. The magnetic field
couples electrostatically to the island through the underlying LL's:
when $B$ increases by $\phinod / A$, the number of electrons in each
of the $\fnod$ underlying LL's will increase by one. These LL's will
act as gates to the isolated island: Coulomb repulsion favors a
constant total electron number inside the device, so $N$ will
decrease by $\fnod$ for every $\phinod / A$ change in $B$, giving
rise to $\fnod$ resistance oscillations. This picture not only
explains the approximate proportionality of $\dB$ to $B$, because $B
\sim 1/\fnod$, but also explains small deviations from it. As seen
in Figs.~2(e) and 2(f), the $\dB$ data is flatter than the
straight-line fit. This picture actually predicts a constant $\dB$
for a given $\fnod$, and the observed increase of $\dB$ can be
accounted for as the device area shrinks slightly at higher fields.

\begin{figure}[!]
\center \label{fig3}
\includegraphics[width=3.10in]{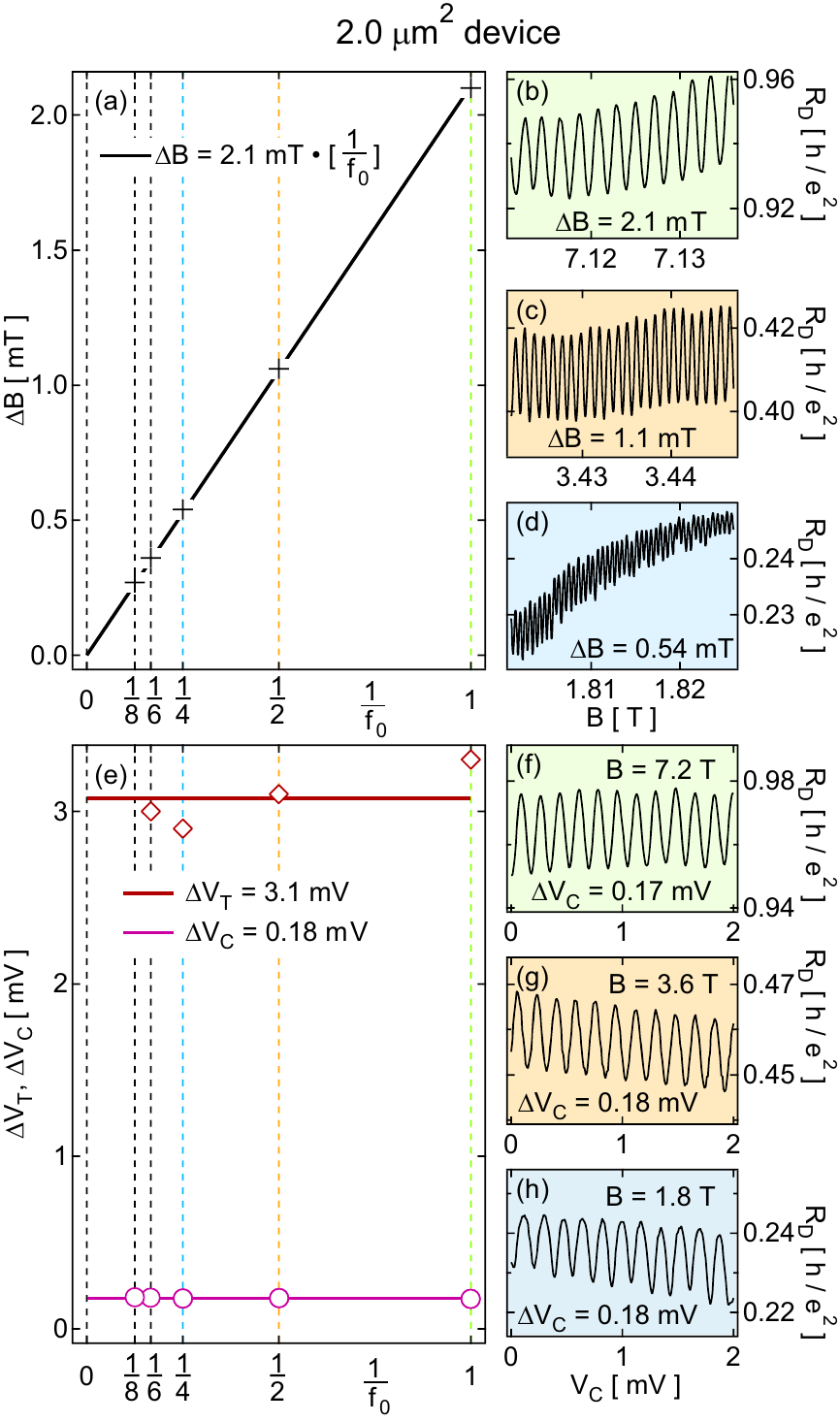}
\caption{\footnotesize{Magnetic field and gate voltage periods at
various $\fnod$, for the $2.0~\squm$ device. (a) $\dB$ as a function
of $1/\fnod$, and a best-fit line constrained through the origin.
(b-d) $\Rd$ oscillations as a function of $B$, at $\fnod=1$, 2, and
4, respectively. (e) $\dVt$ (diamonds) and $\dVc$ (circles) as a
function of $1 / \fnod$, and their averages indicated by horizontal
lines. (f-h) $\Rd$ oscillations as a function of $\Vc$, at
$\fnod=1$, 2, and 4, respectively.}}
\end{figure}

Motivated by this picture, in Fig.~3(a) we show the average $\dB$ at
each $1/\fnod$, and a straight-line fit constrained through the
origin, demonstrating the expected relationship $\dB = (\phinod / A)
/ \fnod$, with $A = 2.0~\squm$. Further evidence of the CB mechanism
in the $2.0~\squm$ device is found in the resistance oscillations as
a function of gate voltages. Figures~3(f-h) show $\Rd$ as a function
of center gate voltage $\Vc$, for $\fnod=1$, 2 and 4, respectively.
Figure~3(e) summarizes gate voltage periods $\dVt$ and $\dVc$ at
various $\fnod$, and shows they are independent of $\fnod$. This
behavior is consistent with the CB mechanism, as gate-voltage
periods are determined by the charging energy and lever arm to the
gate, both of which are approximately independent of $B$.

\begin{figure}[!]
\center \label{fig4}
\includegraphics[width=3.03in]{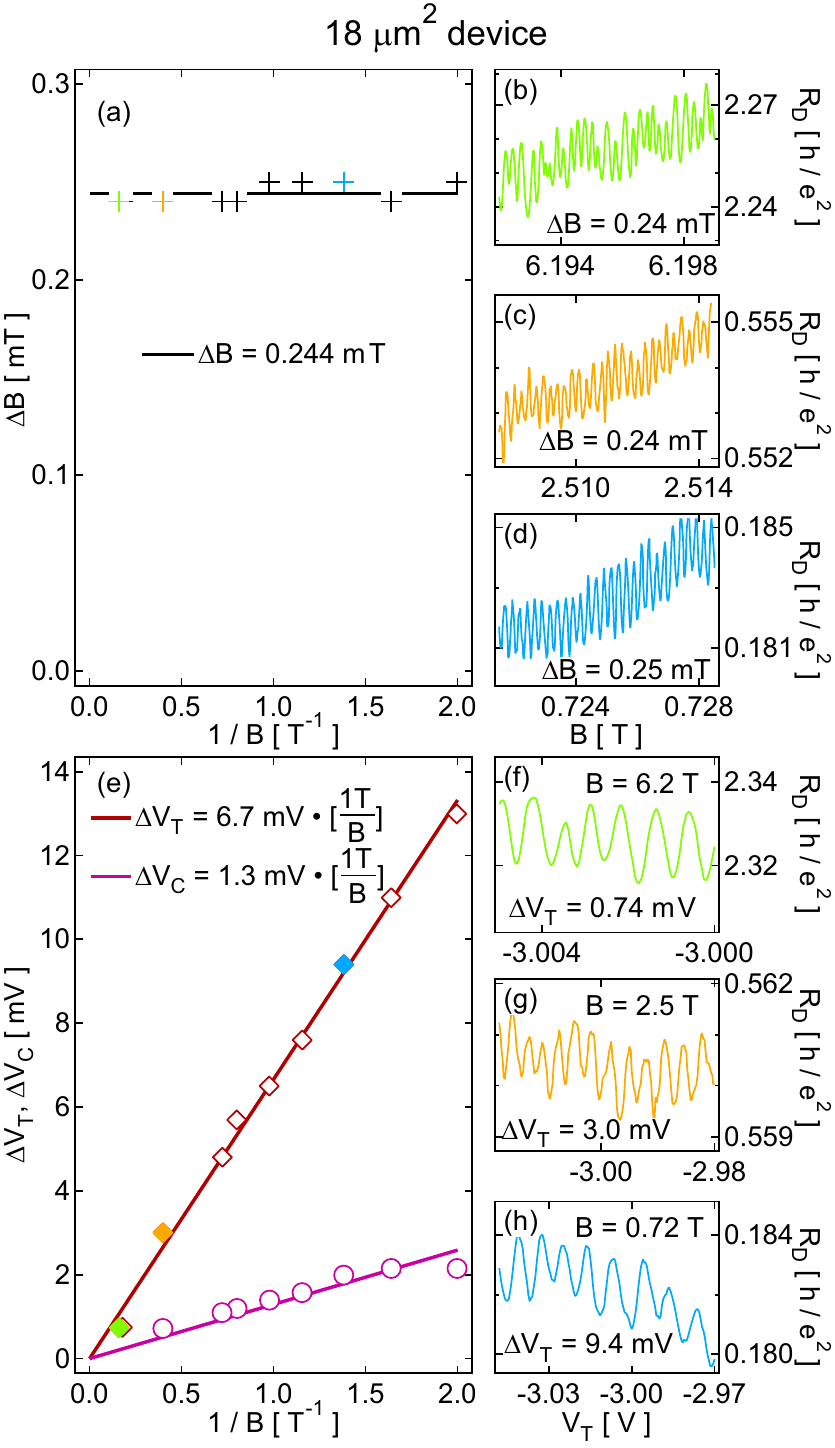}
\caption{\footnotesize{Magnetic field and gate voltage periods at
various $B$, for the $18~\squm$ device. (a) $\dB$ as a function of
$1 / B$, and their average indicated by a horizontal line. (b-d)
$\Rd$ oscillations as a function of $B$, over three magnetic field
ranges. (e) $\dVt$ (diamonds) and $\dVc$ (circles) as a function of
$1 / B$, and best-fit lines constrained through the origin. (f-h)
$\Rd$ oscillations as a function of $\Vt$, at $B = 6.2~\Tesla$,
$2.5~\Tesla$, and $0.72~\Tesla$, respectively. }}
\end{figure}

Having identified CB as the dominant mechanism for resistance oscillations in the
$2.0~\squm$ device, we fabricated and measured an $18~\squm$ device, an order of
magnitude larger in size, hence an order of magnitude smaller in charging energy.
The center gate covering the whole interferometer, not present in previous
experiments~\cite{vanWees89,Alphenaar92CB,Camino07CB,Godfrey07}, also serves to
reduce the charging energy. In this device, $\Rd$ as a function of $B$ at three
different fields is plotted in Figs.~4(b-d), showing nearly constant $\dB$. The
summary of data in Fig.~4(a) shows that $\dB$, measured at 10 different fields
ranging from $0.5$ to $6.2~\mathrm{T}$, is indeed independent of $B$; its average
value of $0.244~\mathrm{mT}$ corresponds to one $\phinod$ through an area of
$17~\squm$, close to the designed area. This is in contrast to the behavior
observed in the $2.0~\squm$ device, and is consistent with AB interference. Gate
voltage periods are also studied, as has been done in the $2.0~\squm$ device.
Figures~4(f-h) show $\Rd$ as a function of $\Vt$ at three different fields, and
Fig.~4(e) shows both $\dVt$ and $\dVc$ as a function of $1/B$. In contrast to the
behavior observed in the $2.0~\squm$ device, $\dVt$ and $\dVc$ are no longer
independent of $B$, but proportional to $1/B$. This behavior is consistent with AB
interference, because the total flux is given by $\phi = B \cdot A$ and the flux
period is always $\phinod$; assuming that the area changes linearly with gate
voltage, gate-voltage periods would scale as $1/B$ for AB. Note that here, the
gate voltage periods can vary smoothly with $B$ and do not correspond to changes
in a quantized charge.

\begin{figure}[!]
\center \label{fig5}
\includegraphics[width=2.40in]{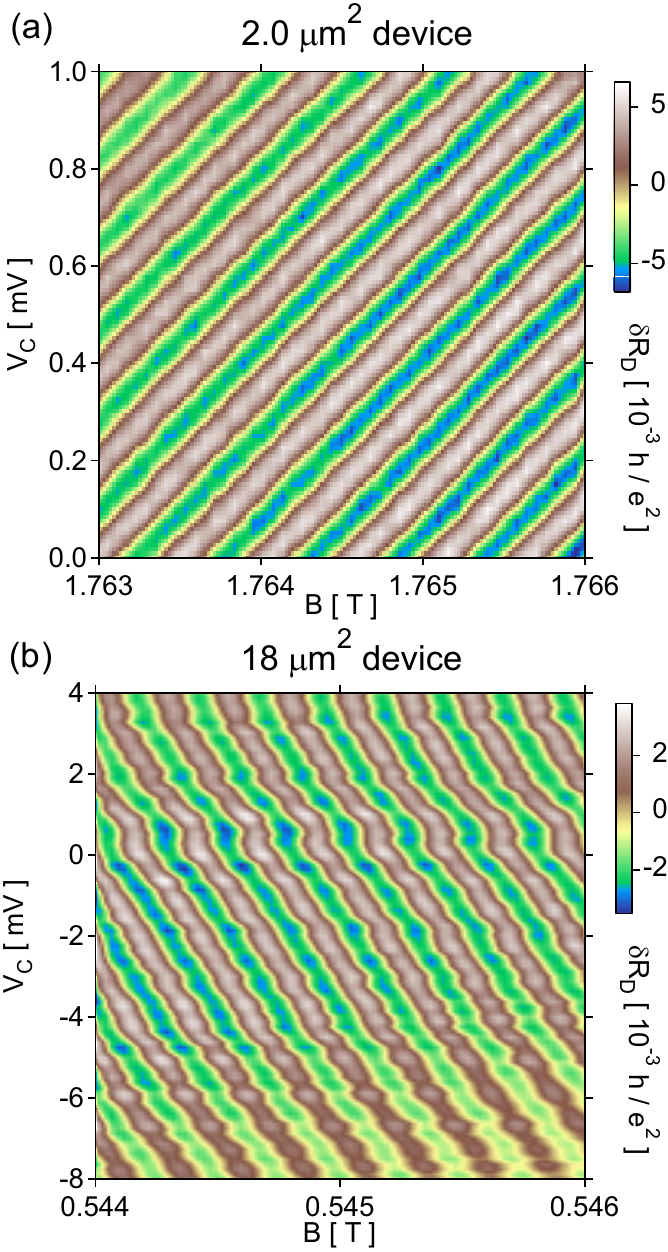}
\caption{\footnotesize{(a) $\dRd$, i.e. $\Rd$ with a smooth
background subtracted, as a function of $B$ and $\Vc$, for the
$2.0~\squm$ device. (b) Same as in (a), but for the $18~\squm$
device.}}
\end{figure}

As shown above, the magnetic field and gate voltage periods
have qualitatively different $B$ dependence in the $2.0~\squm$ and
$18~\squm$ devices, the former consistent with CB, and the latter
consistent with AB interference. Based on these physical pictures,
one can make another prediction in which these two mechanisms will
lead to opposite behaviors. In the CB case, increasing $B$ increases
the electron number in the underlying LL's, thus reducing the
electron number in the isolated island via Coulomb repulsion. This
is equivalent to applying more negative gate voltage to the device.
On the other hand, for the AB case, increasing $B$ increases the
total flux through the interferometer, and applying more positive
gate voltage increases the area, thus also the total flux;
therefore, higher $B$ is equivalent to more positive gate voltage.
As a result, if $\Rd$ is plotted in a plane of gate voltage and $B$,
we expect stripes with a positive slope in the CB case and a
negative slope in the AB case.

Figures~5(a,b) show $\Rd$ as a function of $\Vc$ and $B$ for the
$2.0~\squm$ and $18~\squm$ devices, respectively. As anticipated,
the stripes from the $2.0~\squm$ device have a positive slope,
consistent with the CB mechanism, while stripes from the $18~\squm$
device have a negative slope, consistent with AB interference. This
difference can serve to determine the origin of resistance
oscillations without the need to change magnetic field
significantly.

We gratefully acknowledge J. B. Miller for device fabrication and
discussion, R. Heeres for his work on the cryostat, and I. P. Radu,
M. A. Kastner, B. Rosenow and B. I. Halperin for helpful
discussions. This research is supported by Microsoft
Corporation Project Q, IBM, NSF (DMR-0501796), and Harvard
University.


\end{document}